\documentclass[11pt]{article}
\usepackage{epsfig}
\begin{document}
\def\eg{ {\em e.g.}~}
\centerline{The Nobel Prize in Physics 1999}
The last Nobel Prize of the Millenium in Physics has been awarded jointly 
to Professor Gerardus 't Hooft of the University of Utrecht in Holland and his 
thesis advisor Professor Emeritus Martinus J.G. Veltman of Holland. According 
to the Academy's citation, the Nobel Prize has been awarded for 
'elucidating the quantum structure of electroweak interaction in Physics'. 
It further goes on to say 
that they have placed particle physics theory on a firmer mathematical 
foundation. In this short note, we will try to understand both these
aspects of the award. The work for which they have been awarded the 
Nobel Prize was done in 1971. However, the precise predictions of properties 
of particles that were made possible as a result of their work,  were tested 
to a very high degree of accuracy only in this last decade. 
To understand the full significance of this
Nobel Prize, we will have to summarise briefly the  developement of our 
current theoretical framework  about the basic constituents of matter and the 
forces which hold them together. In fact  the path can be partially traced in 
a chain of Nobel prizes starting from one in  1965 to S. Tomonaga, J. Schwinger and R. Feynman, to the one to S.L. Glashow, A. Salam and S. Weinberg  in 1979, 
and then  to C. Rubia and Simon van der Meer in 1984 ending with the current 
one.
	
	In the article  on `Search for a final theory of matter' in this issue, 
Prof. Ashoke Sen has described the `Standard Model (SM)' of particle physics, 
wherein he has listed all the elementary particles according to the SM. 
These consist of the matter particles: the quarks and leptons along with 
various vector bosons $\gamma, W^\pm, Z^0$ and gluons $g$ which mediate the 
various interactions between them.  Box 1 summarises them here again for 
sake of completeness. As explained in that article, one of the conceptual 
cornerstones of the 
current description of particle physics is the fact that an interaction (say 
Coulomb) between two elementary particles (say electrons) can be understood 
{\it either} (i) 
as the effect of the force field generated by one of them on the other one 
{\it or equivalently} 
(ii) as arising due to an exchange of the carrier of the force (photon 
in this case) between them. The photon is the `quantum' of the 
electromagnetic field. The range as well as the dependence of this force on 
the relative spins  and positions of the particles is correlated with the 
properties of this quantum  and this can be  established in a well defined 
mathematical framework. Box 2 depicts this equivalence in a pictorial manner.

	We know that just as Newtonian mechanics is the right mathematical 
framework to describe the terresterial and celestial motion, quantum mechanics 
is the right language to describe the motion of molecules, atoms, electrons, 
neutrons/protons at the molecular/subatomic and subnuclear level. If we want to 
describe, in addition to these, creations and annihilation of particles, e.g. as 
happens in the spontaneous transitions of an atom, we need to further extend 
this mathematical framework to the next higher level of sophistication called 
`Quantum Field Theory' (QFT). In QFT not only that we employ fields to 
describe the carriers of interaction, the matter particles are also described 
by matter fields.

	Another cornerstone of our theoretical understanding of the fundamental 
particles and their interactions is the realization of the important role played 
by Symmetries / Invariances. The idea of symmetries can be understood in the 
following way. Laws of physics,let us say 
$\vec F = m {{d^2\vec x } \over {dt^2}}$, should be the same no matter which 
point in the universe  do we choose as the origin  of our 
coordinate system. This means that the physics is unchanged under a change of 
the origin of the coordinate system. This is expressed by saying 
that the system is invariant under a transformation of coordinates involving
translations in space. As per our current understanding,  
underlying,fundamental invariance principles actually dictate 
the form of interactions. Let us understand it by taking the example of 
gravitation. Newton deduced the law of gravitation from the observation of 
motion.  On the other hand  Einstein wrote down the general theory of 
relativity by postulating that the description of motion should be the same 
for two observers employing two coordinate systems which are related to each 
other by a  general transformation. In particular the transformation 
can be different at different points in space-time.  The non-relativistic limit 
of this theory (i.e. when objects move at speeds much lower than light) 
contains Newton's theory of gravitation. Thus the `general 
co-ordinate invariance' `explains' the laws of gravitation `deduced' 
by Newton.  So in some sense we have a theoretical
udnerstanding of an observed law of nature in terms of a deeper guiding 
principle.  The tenet of current theoretical description of 
particle physics is that the Quantum Field Theories which have certain 
invariances are the correct theoretical framework for this description.

The invariance that is most relevant for the discussion here is the so called 
`local gauge invariance'. Without going into the details of the idea,
let us just note that this is basically a generalization of the idea that in 
electrostatics the Electric field and hence the electrostatic force depends only on 
the {\it difference} in potential, and not on the actual values of  the 
potential, i.e. setting of the zero of the potential scale is arbitrary as 
far as the force is concerned.

Quantum Field Theories , though now {\it the} accepted framework for describing
particles and their interactions, were in the doghouse for a long time 
in the 60s 
because they used to predict nonsensical, infinite results for properties of 
particles when one tried to compute them accurately. The difficulties arise 
essentially because of the nontrivial structure that the vacuum has in QFT. 
This 
can be visualized by thinking about  effect that a medium has on particle 
properties; \eg the transport of an electron in a solid can be described 
more easily by imagining that its mass gets changed to an `effective' mass.
Another example is the polarization of the charges, in a dielectric 
medium, caused by a charged particle. This polarization can cause a 
`change' of the charge of the particle. In QFT, vacuum acts as a 
nontrivial medium. The troublesome part, however, is that when one tries to 
calculate this  change in the charge due to the `vacuum polarisation', one gets 
infinite results.  Tomonaga, Schwinger  and Feynman (who got 
the Nobel Prize in 1965)  put the Quantum Field Theoretic description of the 
electron and photon (Quantum Electrodynamics) on a firm mathematical footing 
They showed how one can use the theory to make sensible, 
testable predictions for particle properties (such as a small shift in the 
energy level of an electron in the Hydrogen atom due to the effect of 
vacuum polarization), in spite of these infinities. If this can be done 
always in a consistent manner, then the corresponding QFT is 
said to be {\it renormalizable}. The point to note is that the `local gauge 
invariance' mentioned earlier  was absolutely essential for the proof of 
renormalizability of Quantum Electrodynamics (QED).
	
The best example where the predictions of this theory were tested to an 
unprecedented accuracy is the measurement of gyromagnetic ratio of the $e^-$
{\it viz.}
$g_e$. This is predicted to be 2 based on a quantum mechanical equation which 
is written down with the requirement  that the description of the $e^-$ is 
the same for two observers
moving relative to each other with a constant velocity. (Dirac equation for the
cognocsenti). The experimentally measured value is close to 2 but differs from 
it significantly. In QED one can calculate the corrections to the value 
of $g_e=2$ coming from effects of interaction of the electron whereby it emits
a $\gamma$ and absorbs it again, in a systematic fashion. Box 3 indicates some
of these corrections.  The measured value agrees with theoretical prediction 
to 11 significant places as shown in Box 3. 

	Thus to summarize so far, the electromagnetic interactions between the 
electron and photons can be described in terms of a QFT. The 
description has immense predictive power due to the property of 
renormalizability that the theory has.  The theory has this
property only because it of its invariance under a set of transformations 
called U(1) local gauge transformations.  

With this, we come to a point in the history in 1971 when  particle 
physicists had a unified description of electromagnetic and weak interaction in 
terms of exchange of $\gamma, W^\pm$ and $Z^0$. S. Weinberg, A. Salam and S. 
Glashow later got the Nobel Prize in 1979 for putting forward this
EW  model.  Just as unification of electricity and 
magnetism by Maxwell had predicted the velocity of light 'c' in terms of the 
dielectric constant and magnetic permeability $\epsilon_0$ and $\mu_0$  of the 
vacuum, this unification predicted values of masses $M_W, M_Z$ in terms of the 
ratios of two coupling strengths, called $\sin^2 \theta_W$. These coupling 
strengths are the analogue of the electric charge in QED. Details of these
relations are displayed in Box 4. C. Rubia and Simon Van der Meer got the 
Nobel Prize in 1984 for discovering the $W^\pm$ and $Z_0$ bosons with masses
and decays as predicted by the EW model.

The $W^\pm, Z^0$ bosons were found to have nonzero masses 
$(M_W = 80.33 \pm 0.15\; GeV, M_Z = 91.187 \pm 0.007\; GeV $ where 1 $GeV$ is 
approximately the mass of a proton). As a result the early efforts  
to  cast this electroweak model in the framework of a QFT, by using a more 
complicated gauge invariance suggested by a generalisation of QED, 
met with failure. Their nonzero mass makes a QFT incorporating these
bosons noninvariant under these gauge transformations. 
This makes the  theory nonrenormalizable. This means calculating corrections 
to the relation 1 in Box  4 is again riddled with infinities.

	 At around the same time P. Higgs and others had proposed a way
to  write a QFT of  {\it massive} $W^\pm, Z^0$ bosons, where the mass term
did not spoil the gauge invariance of the theory. This required
existence of an additional particle called the Higgs boson. 
This is where 't Hooft and Veltman stepped in. 't Hooft demonstrated, in 
his thesis work and the paper published in Nuclear Physics B in 1971,  
first that the QFT with massless $W^\pm$ and $Z^0$  was renormalizable 
and the
invariance of the theory under more complex noncommutative local gauge 
transformations was  essential for that. He further showed that a QFT 
containing ${\it massive}$  $W^\pm, Z^0$ bosons would be renormalizable 
(i.e., coefficients of infinite corrections would vanish identically) 
{\it inspite} of nonzero masses as long as the mass was generated through the 
mechanism suggested by P. Higgs. Together `t Hooft and Veltman developed 
new methods of calculation for the higher order corrections to particle 
properties, which explicitly preserved this gauge invariance. This 
work opened the floodgates of the prospects of using the ElectroWeak theory to 
make accurate predictions and test the theory to a similar degree of accuracy  
as the QED {\it cf.} Box 3. Veltman led the program of calculation of various
higher order corrections to EW quantities, having established that the 
results were guaranteed to be finite. He actually developed  a computer 
program called `Schoonship' to use the computer to do these very  complicated
analytical calculations specific to Theoretical High Energy Physics. 
This  is the sense in which the work of 
't Hooft and Veltman put the EW theory on a firm mathematical footing. 
This work was enough to  convince the particle theorists that gauge theories 
with Higgs mechanism was the way to go to describe EW interactions.

	In QED the corrections (\eg to $(g-2)_e$ shown in Box 3) depended 
only on the mass and charge of an $e^-$, wheras in EW theory they depend on the free parameters of this theory viz. the 
masses of various quarks and leptons. The corrections are dominated by the top 
quark due to its large mass. Box 4  shows the leading  corrections predicted 
in the EW theory to the ratio $\rho = { {M^2_W} \over {M^2_Z cos^2 \theta_W}}$. 
The measurement of $M_W / M_Z$ and $\sin^2 \theta_W$  in 
1984 were consistent with $\rho =1$ which was the analogue of $g=2$
prediction of QED.  The measurements then were not precise enough to decide 
what the deviation of experimentlly measured value of $\rho$  from 1 was. 
In the decade since then, a detailed study of the properties of these bosons
has been possible using the 10 million $Z^0$ bosons created 
at the Large Electron Positron Collider (LEP) in Geneva and thousands 
of  $W^\pm$ bosons at the $p \bar p$ collider Tevatron at Fermilab in 
Chicago. By 1993 $\rho $ was found to be $1.011 \pm 0.006$. This implied, 
as can be seen from the Box 4, that the top quark, which was not discovered 
till 1995 must have a mass $M_t \sim 180 GeV$. Finding the top quark in 
1995 with a mass consistent with this value  indeed tested the predictions of 
the EW theory to high accuracy.  The precision of these measurements meant that 
if one did not use the corrected expressions, the values of $M^2_W, M^2_Z$
and $\sin^2 \theta_W$ would not be consistent with each other  within the SM.

Even though not shown in Box 4, the corrections to this ratio also depend
on the mass of the {\it  only particle in the SM which is as yet undiscovered} 
viz. the Higgs Boson, albeit very weakly.  The figure in Box 5  shows the 
region in the $M_W - M_t$  plane that is indicated by measurements today. 
The straight line shows predictions of the SM for different 
values of the Higgs boson. So just as five years ago, one used these 
measurements to `determine' values of $M_t$ (which was then not measured) 
now particle physicists are using them to `determine' the mass of the 
elusive Higgs particle. These precision measurements narrow down the mass 
range where the Higgs boson is likely to be found if SM is indeed completely 
correct. Hunt for this will be on at the Large Hadron Collider (LHC)
which will go in action in 2006.

	The EW theory predicts a slew of measurable quantities in terms of the 
basic parameters of the theory viz. the couplings and masses of quarks/leptons. 
Fig. in Box 6 shows a comparison of the predictions of the SM (corrected for 
these loop effects) with data. The numbers in the third column indicate the 
difference between the prediction and measurement in units of the standard 
deviation. It is this agreement, which would be nowhere as excellent if we do not 
include the higher order corrections, that has proved that the EW interactions 
are correctly described in terms of a Quantum Field Theory whose 
renormalizability was established by 't Hooft and Veltman's work. Their 
Nobel prize is also the recognition of the success of QFT and Gauge Principle 
which are the two cornerstones of the mathematical description and 
understanding of the electromagnetic, weak and strong interactions among 
fundamental particles. The  only part of this edifice that is as yet not 
honoured with a Nobel prize is QCD: Gauge theory of strong interactions. 
Who knows, in a few years we will be reading about the work of 
D. Gross, H. Politzer and F. Wilczek in a similar article!
\end{document}